\documentclass[12pt,preprint]{aastex}

\slugcomment{Accepted by ApJ Letters}

\shorttitle{Nonaxisymmetric Collapse in Magnetic Clouds}
\shortauthors{Basu \& Ciolek}

\newcommand{\beq}{\begin{equation}}
\newcommand{\eeq}{\end{equation}}

\newcommand{\cs}{c_{\rm s}}
\newcommand{\cmc}{~{\rm cm}^{-3}}
\newcommand{\cms}{~{\rm cm}^{-2}}

\newcommand{\kms}{~\rm km~s^{-1}}

\newcommand{\pc}{~\rm pc}

\newcommand{\mH}{m_{\rm H}}
\newcommand{\Pe}{P_{\rm ext}}
\newcommand{\mui}{\mu_0}
\newcommand{\muc}{\mu_{\rm core}}
\newcommand{\muenv}{\mu_{\rm env}}

\newcommand{\tnii}{\tilde{\tau}_{\rm ni,0}}
\newcommand{\Pext}{\tilde{P}_{\rm ext}}
\newcommand{\rhon}{\rho_{\rm n}}
\newcommand{\nn}{n_{\rm n}}

\newcommand{\nni}{n_{\rm n,0}}
\newcommand{\nion}{n_{\rm i}}
\newcommand{\sign}{\sigma_{\rm n}}

\newcommand{\signi}{\sigma_{\rm n,0}}
\newcommand{\Beq}{B_{z,\rm eq}}

\newcommand{\Bref}{B_{\rm ref}}
\newcommand{\vn}{v_{\rm n}}
\newcommand{\vnx}{v_{{\rm n},x}}
\newcommand{\zhat}{\mbox{\boldmath$ \hat{z}$}}

\newcommand{\lammax}{\lambda_{\rm T,m}}

\begin{document}

\title{Formation and Collapse of Nonaxisymmetric Protostellar Cores
in Planar Magnetic Molecular Clouds}
\author{Shantanu Basu\altaffilmark{1} and Glenn E. Ciolek\altaffilmark{2}}
\altaffiltext{1}{Department of Physics and Astronomy, University of
Western Ontario, London, Ontario N6A~3K7, Canada; basu@astro.uwo.ca.}
\altaffiltext{2}{New York Center for Studies on the Origins of Life (NSCORT),
and Department of Physics, Applied Physics, and Astronomy,
Rensselaer Polytechnic Institute, 110 W. 8th Street, Troy, NY 12180;
cioleg@rpi.edu.}

\begin{abstract}
We extend our earlier work on ambipolar diffusion induced formation of
protostellar cores in isothermal sheet-like magnetic interstellar
clouds, by studying nonaxisymmetric collapse for the physically
interesting regime of magnetically critical and supercritical model
clouds ($\mui \geq 1$, where $\mui$ is the initial mass-to-magnetic
flux ratio in units of the critical value for gravitational collapse).
Cores that form in model simulations are effectively triaxial, with
shapes that are typically closer to being oblate, rather than prolate. Infall
velocities in the critical model ($\mui = 1$) are subsonic; in contrast,
a supercritical model ($\mui = 2$) has extended supersonic infall that
may be excluded by observations. For the magnetically critical model,
ambipolar diffusion forms cores that are supercritical ($\muc > 1$) and
embedded within subcritical envelopes ($\muenv < 1$).
Cores in our models have
density profiles that eventually merge into a near-uniform background,
which is suggestive of observed properties of cloud cores.
\end{abstract}

\section{Introduction}
Magnetic fields play an important role in star formation, especially in
the early stages of core formation and collapse; measured mass-to-flux
ratios of molecular clouds yield an average that is $\sim 1 - 2$ 
times greater than the critical value for collapse (Crutcher
1999). However, observational biases tend to push toward higher values
of measured mass-to-flux ratio (Crutcher 2003, private communication),
so that moderately subcritical cloud regions are not ruled out.
Dense cores within molecular clouds are the sites of star formation,
with detected infall 
up to $\approx 0.5 \, \cs \approx
0.1 \kms$ on scales $\lesssim 0.1 \pc$, (e.g., in L1544, Tafalla et al.
1998; Williams et al. 1999), where $\cs$ is the isothermal sound speed.
Ciolek \& Basu (2000) have fit the main features of the observed
velocity and density profiles in L1544, modeling it as an axisymmetric
supercritical core embedded in a moderately subcritical envelope.

Axisymmetry is clearly an idealization to real cores. Observations
suggest a typical projected axis ratio of 0.5 for cores (Myers et al.
1991), and deprojections of the distribution of shapes imply
intrinsically triaxial but nearly oblate cores (Jones, Basu,
\& Dubinski 2001). Polarized emission measurements from dense cores also
imply triaxiality (Basu 2000). More generally, detailed submillimeter
maps of star-forming regions reveal significant irregular structure and
multiple cores (Motte, Andr\'e, \& Neri 1998).
Theoretical nonaxisymmetric magnetic models of the collapse and
fragmentation of a single core were presented by Nakamura \& Hanawa
(1997) and Nakamura \& Li (2002), without and with ambipolar diffusion,
respectively, using the magnetic thin-disk approximation (Ciolek \&
Mouschovias 1993, hereafter CM93). The early stages of core formation in
a nonaxisymmetric infinitesimally thin subcritical sheet (including the
effects of magnetic tension but ignoring magnetic pressure) were studied
by Indebetouw \& Zweibel (2001). Here, we also study a planar cloud that
is perpendicular to the mean magnetic field, focusing on the case of
either exactly critical or decidedly supercritical cores; these cases
are shown to lead to observationally distinguishable outcomes. We again
use the thin-disk approximation, which allows for finite thickness
effects, and explicitly includes the effects of both magnetic pressure
and tension.
\vspace{-4ex}
\section{Physical and Numerical Model}
The fundamental equations we use to model molecular clouds as
self-gravitating, partially-ionized, isothermal thin sheets or disks
have been presented in axisymmetric form in CM93 and Basu \& Mouschovias
(1994, hereafter BM94). In this study, the condition of axisymmetry is no
longer employed, and instead we model clouds as thin sheets of infinite
extent [with vertical half-thickness $Z(x,y,t)$] in a cartesian
coordinate system ($x,y$). Magnetohydrostatic equilibrium in the
$z$-direction (i.e., balance of thermal-pressure, gravitational, and
magnetic pinching forces) is assumed at all times. The evolution is
followed in a square region using periodic boundary conditions. To find
the gravitational field components in the sheet, we calculate the
gravitational potential $\psi(x,y,t)$. For this situation, solving
Poisson's equation (see eq. [29] of CM93) relates $\psi$ to the column
density $\sign(x,y,t)$ ($= \int_{-Z}^{Z} \rhon(x,y,t)dz$; $\rhon$ is the
mass density) by
\beq
\label{FTeq}
{\cal F} [\psi] = - 2 \pi G {{\cal F}[\sign]}/k_{z}  ,
\eeq
where ${\cal F}[f]$ is the Fourier transform of a function $f$, $k_{z}$
($= [k_{x}^{2} + k_{y}^{2}]^{1/2}$) is the wavenumber, and $G$ is the
gravitational constant. In our governing equations, the effects of
magnetic pressure and magnetic tension are both included. The magnetic
potential $\Psi(x,y,t)$ that is used to determine the $x$- and $y$-
components of the magnetic field at the surface of the sheet (necessary
to calculate magnetic tension forces) is also given by equation
(\ref{FTeq}), by substituting $\psi$ with $\Psi$, and $2 \pi G \sign$
with $-(\Beq - \Bref)$; for details see CM93.
$\Beq(x,y,t)$ is the vertical magnetic field strength in the sheet, and
$\Bref$ is the uniform magnetic field of the background (reference)
state, as well as the assumed constant magnetic field as
$|z| \rightarrow \infty$.
In the present study, we neglect magnetic braking due to a finite
density external medium (BM94), and dust grains and UV ionization (CM93,
Ciolek \& Mouschovias 1995).

In our formulation, velocities are normalized to the isothermal speed of
sound $\cs = 0.188 \, (T/10~{\rm K})^{1/2}\kms$ (where $T$ is the
temperature and we have used a mean molecular mass $m = 2.33 \mH$, in
which $\mH$ is the hydrogen atom mass), column densities are in units of
$\signi$, the uniform column density of the background state. The time
unit is $t_0 = \cs/2\pi G \signi$, the length unit is
$L_0 = \cs^{2}/2 \pi G \signi$, and the magnetic field unit is
$B_0 = 2 \pi G^{1/2} \signi$.
As discussed in CM93 and BM94, model clouds are distinguished by
three basic dimensionless parameters, namely, the initial mass-to-magnetic
flux ratio (in units of the critical value for gravitational collapse)
$\mui \equiv B_0/\Bref = 2 \pi G^{1/2} \signi/\Bref$, the initial
dimensionless neutral-ion collision time $\tnii \equiv \tau_{\rm ni}/t_0$, and
the normalized external pressure acting on the disk, $\Pext \equiv
\Pe/(\frac{\pi}{2} G \signi^2)$. The ion number density $\nion$ is
calculated using the scaling law $\nion \propto \nn^{1/2}$ (as done by
BM94; $\nn$ is the neutral number density), which is a reasonable
approximation for $\nn \lesssim 10^{6} \cmc$ (e.g., Ciolek \& Mouschovias
1998). For a background number density $\nni = 3 \times 10^3 \, \cmc$
and $\Pext = 0.1$, we find that $\signi = 5.98 \times 10^{-3}$ g cm$^{-2}$
(therefore $N_{\rm n,0} \equiv \signi/m = 1.54 \times 10^{21} \cms$),
$L_0 = 4.57 \times 10^{-2}$ pc, and $t_0 = 2.38 \times 10^5$ yr.

We use the method of lines technique (e.g., Schiesser 1991); the
system of partial differential equations of two-fluid MHD are converted
to a system of ordinary differential equations (ODE's) through second-order
spatial finite differencing, and use of the van Leer (1977) advection scheme.
Time integration of the ODE's is performed using the
implicit Adams-Bashforth-Moulton method. Details of
our basic technique can be found in Morton, Mouschovias, \& Ciolek
(1994). Fast Fourier transform (FFT) routines are used to solve for the
gravitational and magnetic potentials obtained through equation
(\ref{FTeq}). The computational region has extent
$4 \lammax$ on each side, where $\lammax=4 \pi L_0$ is
the wavelength with maximum growth rate of gravitational instability
in a non-magnetic (thermal) infinitesimally thin disk.
Models presented in this paper are run on a uniform grid
of $128 \times 128$ points.
\vspace{-4ex}
\section{Results}
We present the evolution of two model clouds;
one model is exactly critical,
with $\mui =1$, and the other is supercritical with $\mui =2$.
Both models have $\tnii = 0.16$ (as adopted in the standard model of
BM94) and $\Pext = 0.1$.
Since the background
(reference) state is characterized by a uniform column density $\signi$ and
magnetic field $\Bref \zhat$ (where $\Bref = B_0/\mui$)
the gravitational and magnetic forces in the reference state
are identically zero.
To initiate evolution, we superimpose
a set of random, small-amplitude (the rms is 3\% of the background)
column density perturbations $\delta \sign$. The magnetic field is
also
perturbed, with $\delta \Beq \propto \delta \sign$, 
to maintain flux-freezing in the initial state of each model. The
spectrum of perturbations is flat (white noise), with damping so that
wavelengths of twice the grid spacing and smaller are negligible; we
choose this particular type of spectrum so as not to preferentially
excite modes with the maximum growth rate.
The evolution of each model is followed until the maximum
column density $\approx 10 \, \signi$. Beyond this
enhancement, gravitational instability cannot be spatially resolved, and
the evolution is relatively very rapid in a small region near the
density peaks, making a simulation of the larger cloud impractical even
at higher resolution. Since the vertical balance along field lines is
primarily between gas pressure $\rhon \, \cs^2$ and self-gravitational
pressure $\frac{\pi}{2} G \sign^2$, the density has increased by
a factor $\approx 100$.

\subsection{Critical model ($\mui = 1$)}
Figure 1 exhibits
color images of the normalized column density $\sign(x,y)/\signi$ and
the local mass-to-flux ratio $\mu(x,y) = 2 \pi G^{1/2} \sign/\Beq$
over the entire computational region for the critical model
at its final output time, $t = 133.9 \, t_0$.
The time is rather long due to the very
small-amplitude perturbations and because magnetic forces exactly balance
gravity in the flux-freezing limit, requiring instability to
be initiated by random ion-neutral drifts.
Two major density peaks have formed by this time. The column density
structure shows that the lowest density contours are the most
irregular and elongated. Henceforth,
we define a dense core to be the region within the $\sign/\signi = 2$ contour,
corresponding to the observationally significant threshold
$\nn \approx 10^4 \cmc$. The masses of the two dense cores in
this simulation are $5.3 M_{\odot}$ (lower left of Fig. $1a$)
and $4.2 M_{\odot}$ (upper right of Fig. $1a$), respectively, using
the standard values of units given in \S\ 2.
For comparison, the enclosed mass in our periodic computational domain
is $151 M_{\odot}$.
The shape of the cores are generally nonaxisymmetric, a result that was
first seen in the magnetically subcritical planar ambipolar diffusion
models of Indebetouw \& Zweibel (2000).
The mean value of $Z \equiv \sign/(2 \rhon)$ implies one oblate core
with relative axis lengths 0.15:1:1, and another that is triaxial with
relative lengths 0.1:0.7:1, making it more nearly oblate than prolate.
The cores formed in this model always have their shortest axis aligned
normal to the sheet (parallel to the mean magnetic field). Although one
core is effectively oblate in this particular model realization, our
simulations reveal they are most commonly triaxial and more nearly
oblate than prolate, in accordance with statistical analyses of observed
cores (Jones et al. 2001).

Velocity vectors of the neutrals are also displayed in Figure $1a$.
Calculation of the rms speeds of neutrals ($v_{\rm n,rms}$) and ions
($v_{\rm i,rms}$) in the entire computational region yield
$v_{\rm n,rms} = 0.114 \, \cs = 1.14 \, v_{\rm i,rms}$; ions lag
behind neutrals because the motions are ultimately
{\em gravitationally driven}. Furthermore,
{\em all of the infall speeds in this model are subsonic}. The maximum
infall speed is $|\vn| = 0.54 \, \cs$, and speeds of this order occur only
within the cores. As we shall see in \S~3.2, subsonic infall is a
{\em distinct} feature of ambipolar-diffusion initiated collapse in clouds
with $\mui \lesssim 1$. This trend is consistent with multi-line studies
that find subsonic gas speeds in dense protostellar cores (Tafalla et
al. 1998; Williams et al. 1999), and with axisymmetric models of core
formation in subcritical clouds (Ciolek \& Basu 2000).
\begin{figure}
\plottwo{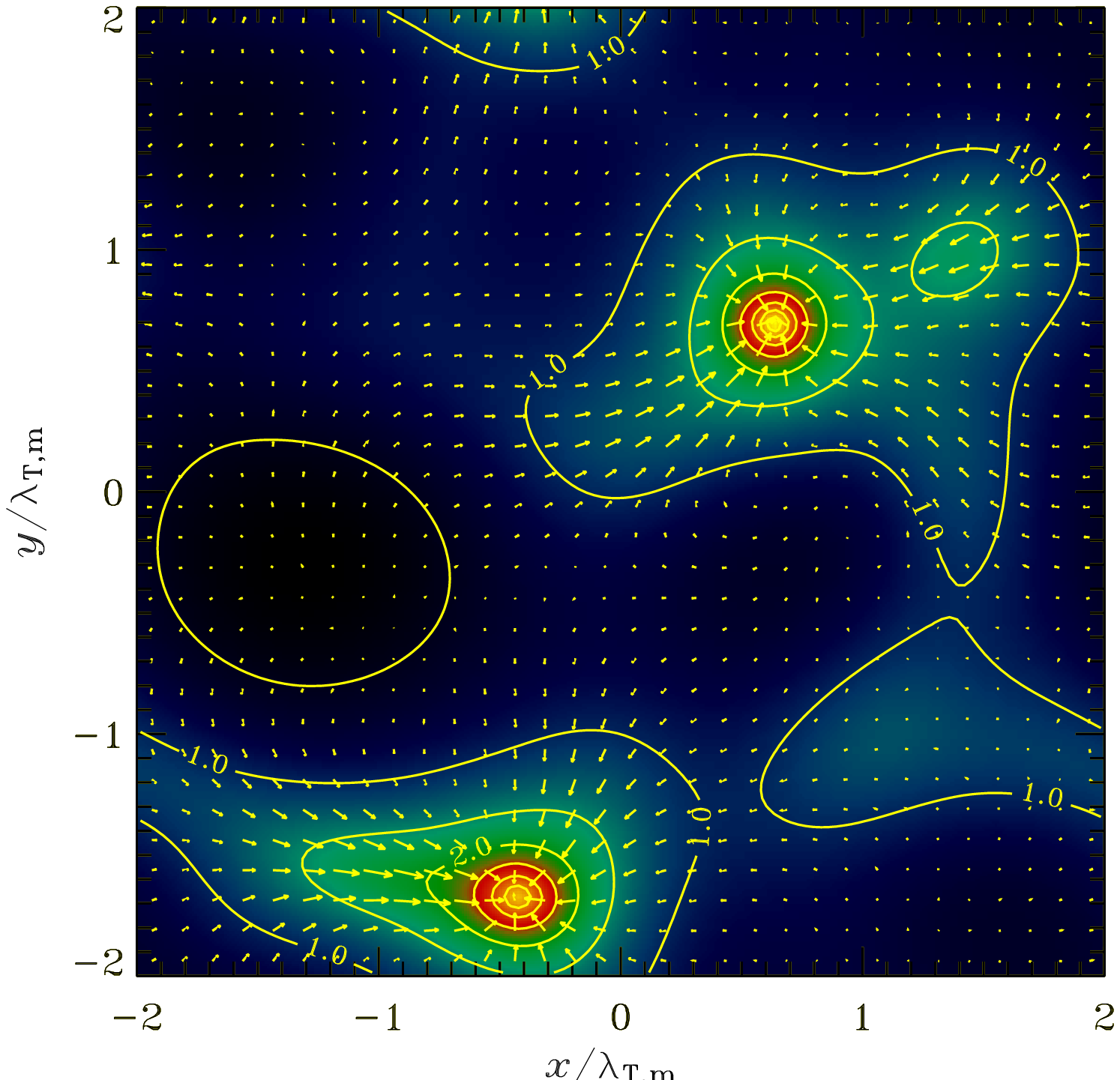}{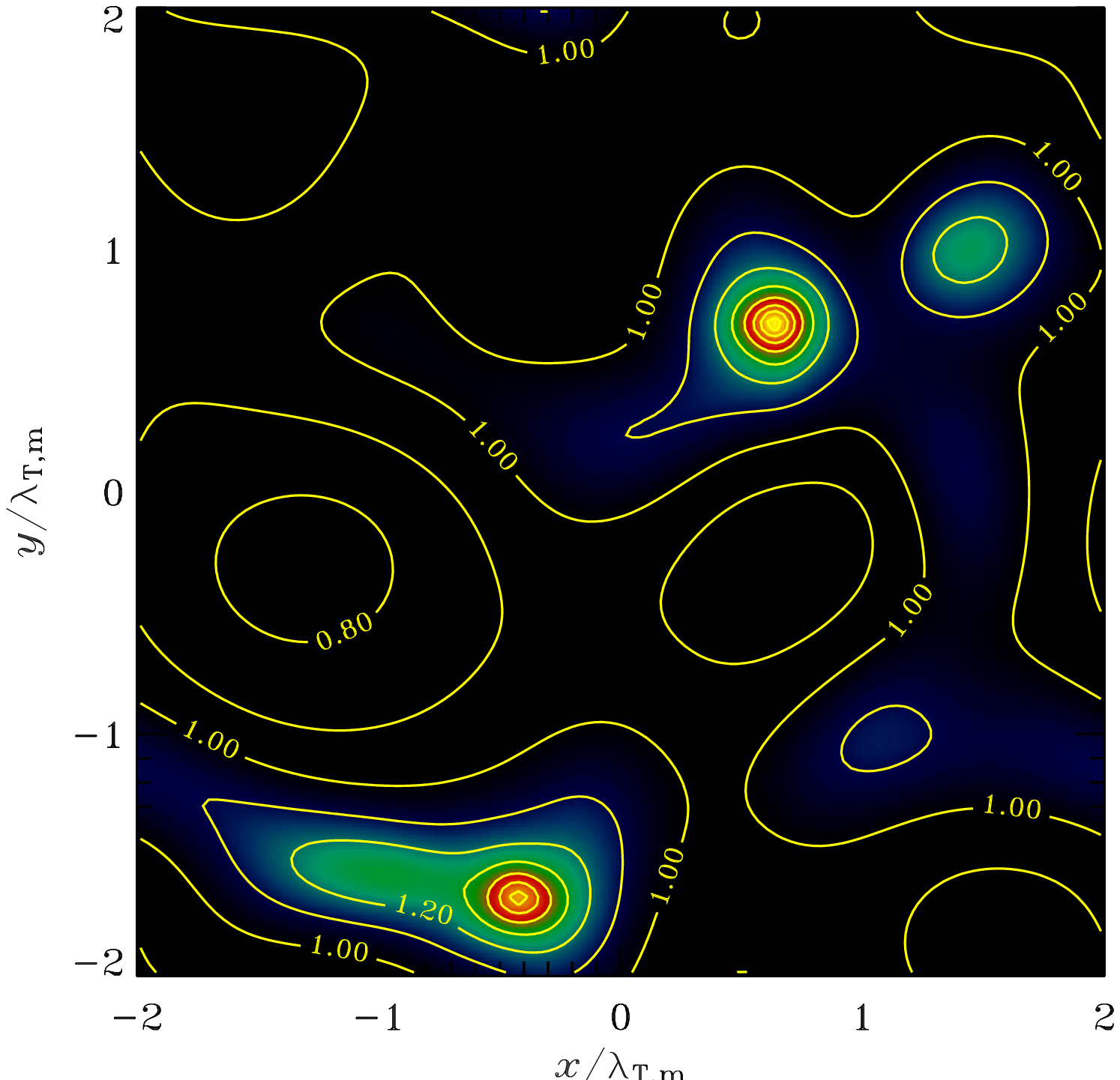}
\caption{Critical model ($\mui = 1$). The data are shown when the maximum
column density $\approx 10 \, \signi$. ({\it a, Left}.) Image and
contour plot of normalized column density $\sign(x,y)/\signi$. The
contour lines are spaced in multiplicative increments of $2^{1/2}$. Also
shown are velocity vectors of the neutrals; the distance between tips of
vectors corresponds to a speed $0.5 \, \cs$. ({\it b, Right.}) Image and
contours of $\mu(x,y)$, the mass-to-flux ratio in units of the critical
value for collapse. Regions with $\mu < 1$ are black. The contour lines
are spaced in additive increments of 0.1.}
\end{figure}
Figure $1b$ shows another interesting result: {\em the cores are
supercritical ($\muc > 1$) and are surrounded by magnetically subcritical
envelopes ($\muenv < 1$)}. This is a natural consequence of ambipolar
diffusion, which redistributes mass in magnetic flux tubes (Mouschovias 1978).
Due to the initial precise balance between gravitational and magnetic
forces in this model, evolution occurs as ambipolar diffusion effects a
drift of mass through essentially stationary magnetic field lines. In
time, this leads to the formation of both supercritical cores
and a subcritical envelope. This can also occur in clouds with 
$\mui$ 
slightly above unity.

Figure 2 shows the column density $\sign/\signi$, vertical magnetic
field $\Beq/B_0$, and $x$-component of neutral velocity $\vnx/\cs$
along the $x$-axis for a line that cuts through one of the cores shown
in Figure 1. The core stands
out as a well-defined high density region with an eventual merger into a
lower-density background (the remnant of the initial uniform state) that
surrounds it; this is reminiscent of the mid-infrared
absorption maps of dense cores made by Bacmann et al. (2000).
The magnitude of $\vnx$ increases inward toward the core
center (but remains subsonic throughout) before dropping to zero at
the core center.
\begin{figure}
\epsscale{0.5}
\plotone{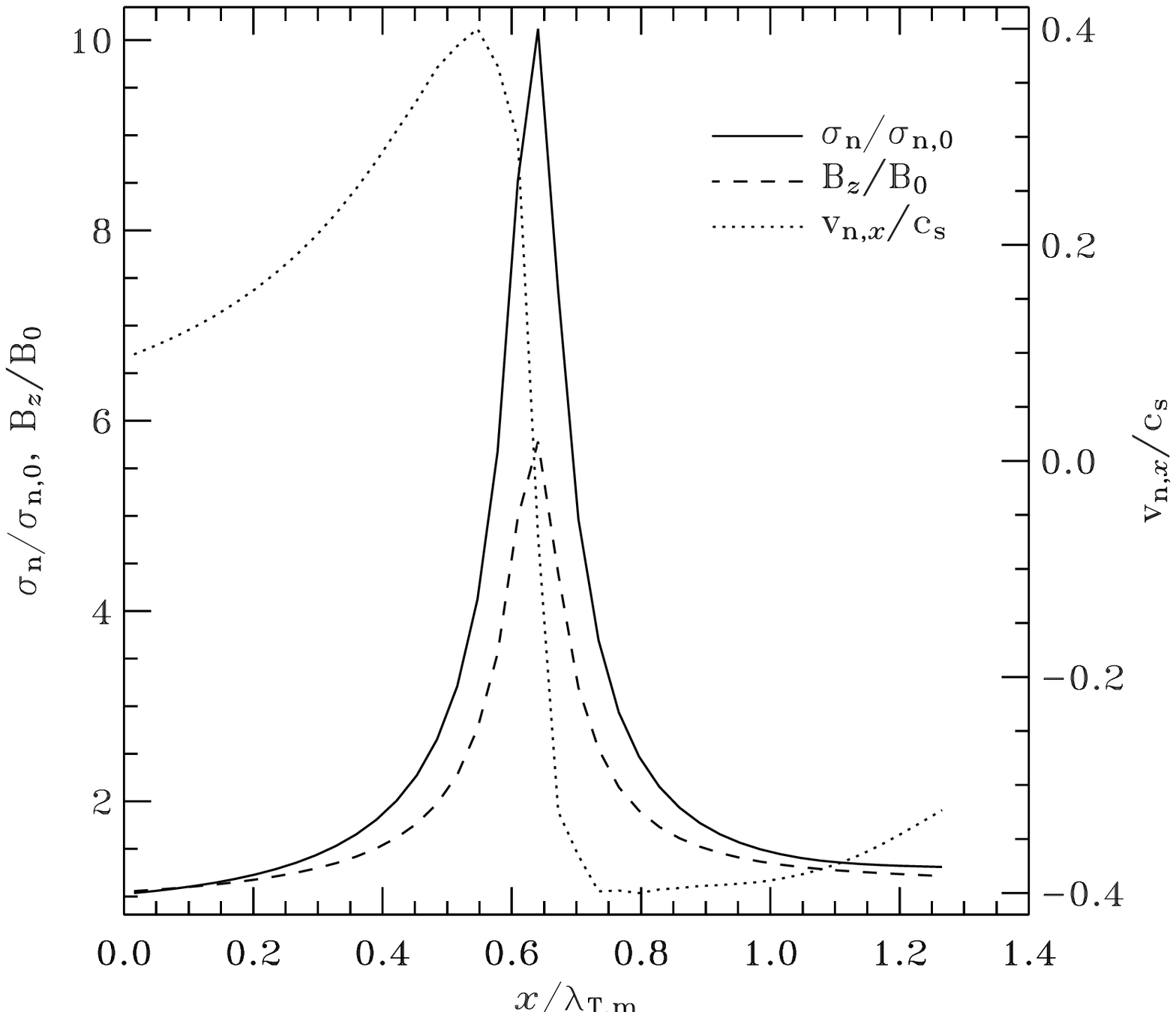}
\caption{Physical quantities in the $\mui = 1$ model, along a line
parallel to the $x$-axis that cuts through a supercritical core (shown
in Figure 1) centered at $x = 0.64 \, \lammax, y = 0.70 \, \lammax$.
Solid curve: column density $\sign/\signi$. Dashed curve: magnetic field
strength $\Beq/B_0$. Dotted curve: $x$-component of the neutral
velocity, $\vnx/\cs$.}
\end{figure}
\subsection{Supercritical model ($\mui = 2$)}
Figure $3a$ exhibits a color image and contour plot of $\sign/\signi$,
as well as velocity vectors for the supercritical model at
its final output time, $t = 17.6 \, t_0$.
The time is significantly lesser than in
the previous model due to the relative ease of gravitational instability
in a supercritical cloud. Because the evolution is more rapid and
dynamical in the supercritical model, ambipolar diffusion doesn't have
as much time to operate.
Hence, the mass-to-flux ratios of the high-density
cores in this model ($\muc \simeq 2.1$) are only slightly greater than
the initial value. The core shapes are again nonaxisymmetric,
either near-oblate or decidedly triaxial.
For this model we find $v_{\rm n,rms} = 0.222 \, \cs = 1.05 \, v_{\rm i,rms}$;
these results differ from the critical model due to the more dynamical
evolution and greater ability of neutrals to drag ions and magnetic
field with them. The maximum infall speed is $1.19 \, \cs$, and speeds of
this order typically occur within distances $\approx 0.1$ pc from the core
centers. In contrast to the critical model, {\it the infall speeds
within the cores in this model are often supersonic, and significant
motions $\approx 0.5 \, \cs$ are also seen throughout the
cloud, and well outside the cores}. Figure $3b$ shows $\sign/\signi$,
$\Beq/B_0$, and $\vnx/\cs$ along the $x$-axis for a line passing
through one of the cores in Figure $3a$.
The collapsing core again has a density profile that eventually
merges into the background.

Our model shows that the extent of the supersonic flow region is well
within the resolution capability of current observations (scales
$\lesssim 0.1~{\rm pc}$, $\nn \approx 10^{4}-10^{5}\cmc$). Since
supersonic infall has not been detected over these length scales and
densities, and models with $\mui \gtrsim 2$ would have even greater
infall speeds, this suggests that {\em molecular clouds that are
supercritical by a factor $\gtrsim 2$ are incompatible with 
observations of protostellar cores.}
\begin{figure}
\epsscale{1.0}
\plottwo{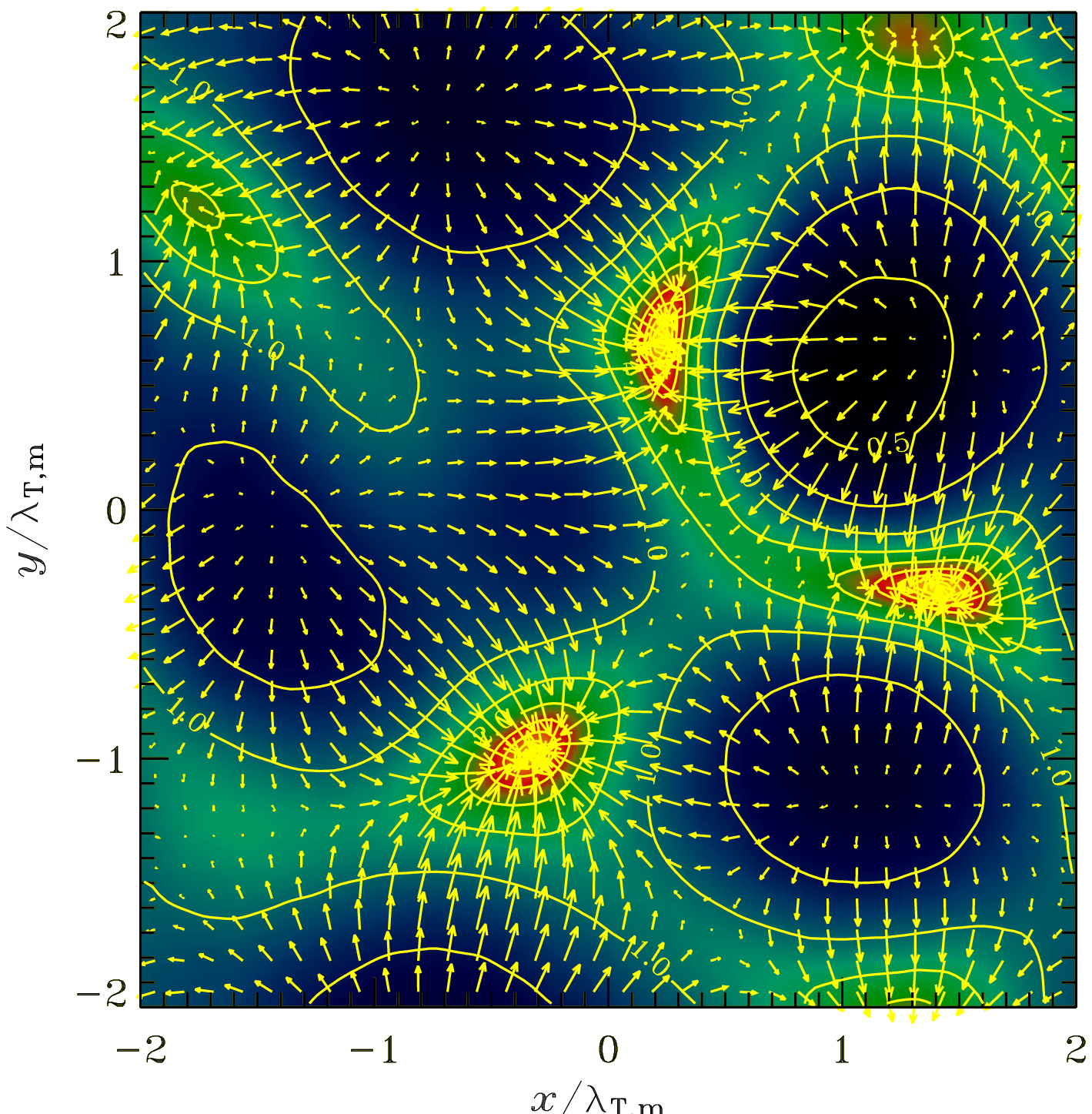}{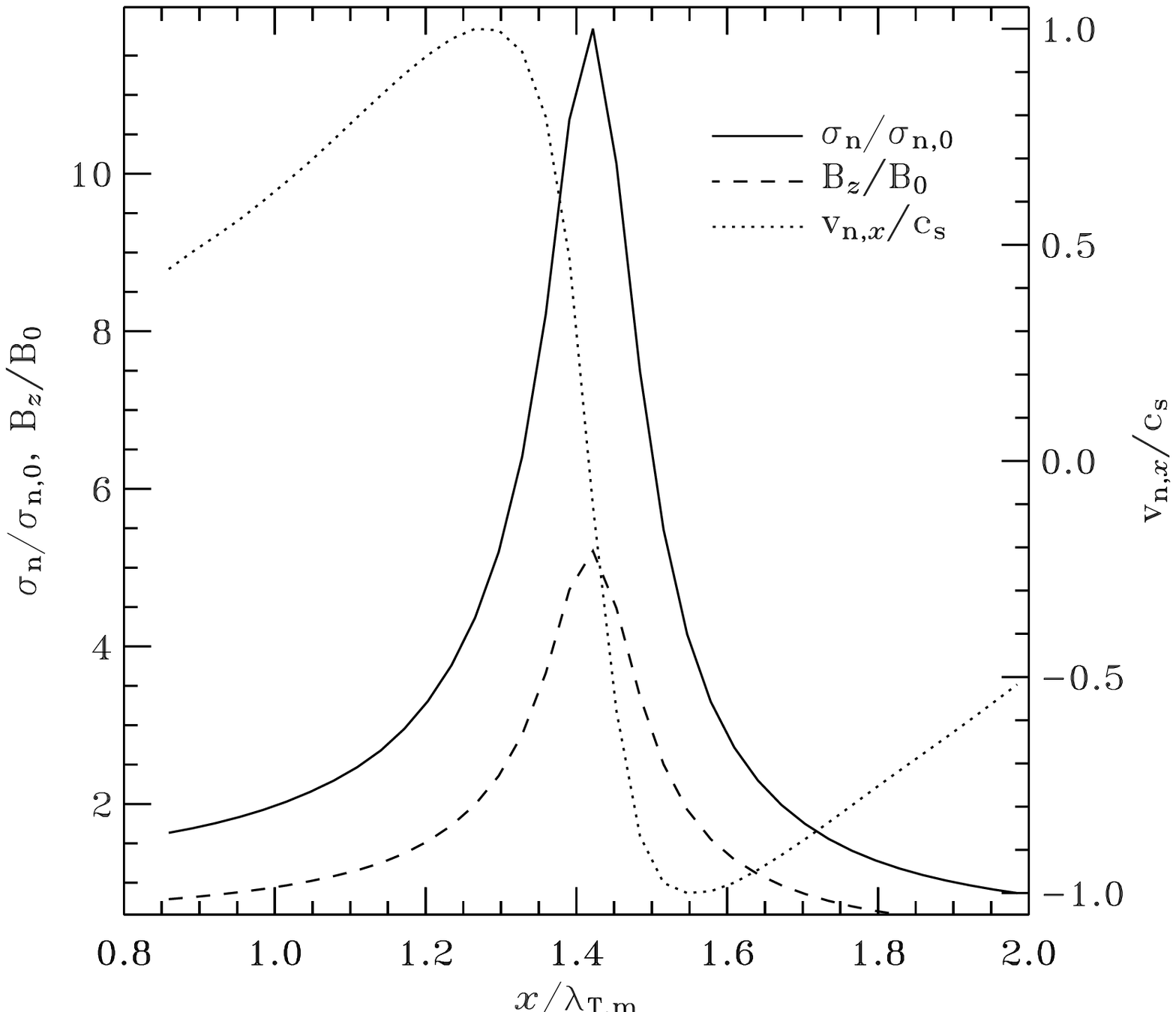}
\caption{Supercritical model ($\mui = 2$). The data are shown when the
maximum column density $\approx 10 \, \signi$. ({\it a, Left}.) Image
and contour plot of normalized column density $\sign(x,y)/\signi$. The
contour lines are spaced in multiplicative increments of $2^{1/2}$. Also
shown are velocity vectors of the neutrals; they are normalized in the same
way as in Fig. $1a$. ({\it b, Right.}) Physical
quantities along a line parallel to the $x$-axis that cuts through a
supercritical core centered at
$x = 1.42 \, \lammax, y = -0.33 \, \lammax$. Solid curve: column density
$\sign/\signi$. Dashed curve: magnetic field strength $\Beq/B_0$.
Dotted curve: $x$-component of the neutral velocity, $\vnx/\cs$.}
\end{figure}

\section{Conclusions}
Ambipolar diffusion leads to a nonuniform distribution of mass-to-flux
ratio, in a natural extension of the process described by Mouschovias
(1978). Stars form preferentially in the most supercritical regions.
Core shapes are somewhat triaxial, and usually
more nearly oblate than prolate. The core column density
eventually merges into a near-uniform
background value. In the critical ($\mui = 1$) model, a surrounding region is
established which is mildly magnetically subcritical
(due to flux redistribution) and infall motions both inside and
outside cores are subsonic. Conversely, the
supercritical ($\mui = 2$) model exhibits supersonic motions within cores, and
extended rapid motions outside them. The critical model requires a
significantly longer time to develop gravitational instability; however,
we caution that the growth time for both models are likely upper limits
due to the possibility of nonlinear perturbations in more realistic
situations. We also note that all motions in these models are
fundamentally gravitationally driven; the neutral speeds are
typically greater than those of the ions.


\begin{acknowledgements}
We thank Craig Markwardt for providing some key IDL routines.
SB is supported by a grant from the Natural Sciences and Engineering
Research Council of Canada. GC is supported by NASA grant
NAG 5-7589 to the New York Origins of Life Center (NSCORT).
\end{acknowledgements}


\begin{thebibliography}{}

\bibitem[]{Bac00} Bacmann, A., Andr\'{e}, P., Puget, J.-L., Abergel, A.,
Bontemps, S., \& Ward-Thompson, D. 2000, \aap, 361, 555
\bibitem[]{B00}  Basu, S. 2000, \apjl, 540, L103
\bibitem[]{BM94} Basu, S., \& Mouschovias, T. Ch. 1994, \apj, 432, 720 (BM94)
\bibitem[]{CB00} Ciolek, G. E., \& Basu, S. 2000, \apj, 529, 925
\bibitem[]{CM93} Ciolek, G. E., \& Mouschovias, T. Ch. 1993, \apj, 454,
194 (CM93)
\bibitem[]{CM95} Ciolek, G. E., \& Mouschovias, T. Ch. 1995, \apj, 418, 774
\bibitem[]{CM98} Ciolek, G. E., \& Mouschovias, T. Ch. 1998, \apj, 504, 280
\bibitem[]{C99} Crutcher, R. M. 1999, \apj, 520, 706
\bibitem[]{IZ00} Indebetouw, R., \& Zweibel, E. G. 2000, \apj, 532, 361
\bibitem[]{JBD01} Jones, C. E., Basu, S., \& Dubinski, J. 2001, \apj, 551, 387
\bibitem[]{MMC94} Morton, S. A., Mouschovias, T. Ch., \& Ciolek, G. E. 1994,
\apj, 421, 561
\bibitem[]{MAN98} Motte, F., Andr\'{e}, P., \& Neri, R. 1998, \aap, 336, 150
\bibitem[]{M78} Mouschovias, T. Ch. 1978, in Protostars \& Planets, ed. T.
Gehrels (Tucson: Univ. Arizona), 209
\bibitem[]{M91} Myers, P. C., Fuller, G. A., Goodman, A. A., \& Benson,
P. J. 1991, \apj, 376, 561
\bibitem[]{NH97} Nakamura, F., \& Hanawa, T. 1997, \apj, 480, 701
\bibitem[]{NL02} Nakamura, F., \& Li, Z.-Y. 2002, \apj, 566, L101
\bibitem[]{Sch91} Schiesser, W. E. 1991, The Numerical Method of Lines:
Method of Integration of Partial Differential Equations (San Diego:
Academic)
\bibitem[]{Ta98} Tafalla, M., Mardones, D., Myers, P. C., Caselli, P.,
Bachiller, R., \& Benson, P. J. 1998, \apj, 504, 900
\bibitem[]{van77} van Leer, B. 1977, JCP, 23, 276
\bibitem[]{Wil99} Williams, J. P., Myers, P. C., Wilner, D. J.,
\& DiFrancesco, J. 1999, \apj, 513, L61

\end{thebibliography}
\end{document}